\documentclass[journal]{IEEEtran}
\usepackage{graphicx}
\usepackage{subfigure}
\usepackage[cmex10]{amsmath}
\usepackage{cite}

\usepackage{hyperref}
\usepackage{multirow}
\usepackage[all]{hypcap} 

\begin{document}
\title{Applications of Many-Core Technologies to On-line Event Reconstruction in High Energy Physics
Experiments}
\author{A.~Gianelle, 
  S.~Amerio, 
  D.~Bastieri, 
  M.~Corvo, 
  W.~Ketchum,
  T.~Liu, 
  A.~Lonardo, 
  D.~Lucchesi,
  S.~Poprocki, 
  R.~Rivera, 
  L.~Tosoratto,
  P.~Vicini
  and 
  P.~Wittich
\thanks{Manuscript received November 22, 2013.
This work was supported by the  US National Science Foundation, the US 
Department of Energy Office of Science and the Italian
Istituto Nazionale di Fisica Nucleare. This work was partially supported by the 
EU Framework Programme 7 project EURETILE under grant number 247846.}
\thanks{D.~Bastieri and D.~Lucchesi are with University of Padova and INFN Padova.}%
\thanks{A.~Gianelle is with INFN Padova.}%
\thanks{M.~Corvo is with University of Ferrara.}%
\thanks{S.~Amerio is with University of Padova.}%
\thanks{W.~Ketchum is with Los Alamos National Laboratory.}%
\thanks{T.~Liu and R.~Rivera are with Fermi National Accelerator Laboratory.}%
\thanks{A.~Lonardo, L.~Tosoratto and P.~Vicini are with INFN Roma.}%
\thanks{S.~Poprocki and P.~Wittich are with Cornell University.}
}
\maketitle
\thispagestyle{empty}
\begin{abstract}
  Interest in many-core architectures applied to real time selections is 
  growing in High Energy Physics (HEP) experiments. In this paper we describe
  performance measurements of many-core devices 
  when applied to a typical HEP online task: 
  the selection of events based on the trajectories of charged particles.
  We use as benchmark a scaled-up version of the algorithm used at CDF 
  experiment at Tevatron for online track reconstruction - the SVT algorithm - 
  as a realistic  test-case for low-latency trigger systems using new computing
  architectures for LHC experiment. We examine the
  complexity/performance trade-off in porting existing serial
  algorithms to many-core devices. We measure performance of
different architectures (Intel Xeon Phi and \textsc{amd} GPUs, in addition 
to NVidia GPUs) and different software
environments (OpenCL, in addition to NVidia CUDA).
Measurements of both data processing  and data transfer latency are shown, 
considering different I/O strategies 
  to/from the many-core devices. \end{abstract}

\section{Introduction}
\IEEEPARstart{R}{eal}-time event reconstruction plays a fundamental role in High Energy
Physics (HEP) experiments at hadron colliders.  Reducing the rate of
data to be saved on tape is critical. To increase the purity of the collected samples,
rate reduction has to be coupled with an initial selection of the most
interesting events.  In a typical hadron collider experiment, the
event rate has to be reduced from tens of MHz to a few kHz.  The
selection system (trigger) is usually organized in successive levels,
each capable of performing a finer selection on more complex physics
objects describing the event. Trigger systems usually comprise a first
level based on custom hardware, followed by one or two levels usually
based on farms of general purpose processors.  At all levels, latency
is a concern: for a fixed processing time, the faster a decision is
rendered about accepting or rejecting an event improves the purity of
the collected data sample. The possibility of using commercial devices
at a low trigger level is very appealing: they are subject to
continuous performance improvements driven by the consumer market, are
less expensive than dedicated hardware, and are easier to support.
Among the commercial devices, many-core architectures such as Graphic
Processing Units (GPUs)~\cite{bib_gpu} and Intel Many Integrated Core
(MIC)~\cite{bib_intelMIC} are of particular interest for online
selections given their great computing power: the latest
\textsc{nvidia}~\cite{bib_nvidia} GPU architecture, Kepler, exceeds
Teraflop computing power. Moreover, high-level programming
architectures based on \textsc{c/c++} such as
\textsc{cuda}~\cite{bib_cuda} and \textsc{opencl}~\cite{bib_opencl}
make programming these devices more accessible to the general
physicist user.  The goal of this study is to investigate the
strengths and weaknesses of many-core devices when applied in a low
latency environment, with particular emphasis on the data transfer
latency to/from the device and the algorithm latency for processing on
the device in a manner similar to a typical HEP trigger application,
and to understand the cost/complexity ratio of porting legacy serial
code to many-core devices.

We showed initial studies on GPU performance in low-latency
environments ($\approx 100~\mu$s) in previous
papers~\cite{TIPP2011,NSS2012,CHEP2013}.  In this paper we extend
those studies to include other many-core architectures (Intel MIC 
and \textsc{amd} GPUs in
addition to \textsc{nvidia} GPUs) and other programming 
toolsets (\textsc{opencl} in addition to \textsc{cuda}).
The algorithm run on the parallel architecture is a complete version
of the fast track-fitting algorithm of the Silicon Vertex Tracker
(SVT) system at CDF~\cite{SVT1}.  
Starting with a serial algorithm implemented on a CPU, 
we test an \textit{embarrassingly
  parallel} algorithm on the Intel MIC environment. In this case each
event is handled independently by a core on the accelerator, and the
parallelization is achieved with only minor changes to the legacy
code.  This approach is only possible in the Intel MIC
environment. Next we consider an algorithm where we unroll three
internal nested loops and run these in parallel on a GPU, using the
\textsc{cuda} and \textsc{opencl} environments. This second approach
is programmatically more complicated and less trivial to implement. In
neither case have we re-thought the basic algorithms or the data
structures used. To achieve optimal performance, these steps would
have to be taken.  As one might expect, the improvement from the first
approach is rather modest, albeit easier to implement, and the second
approach shows larger performance gains. For GPUs, we also test
different strategies to transfer data to and from the device.

\section{SVT track fitting algorithm}
The Silicon Vertex Trigger (SVT)~\cite{SVT1,SVT2} is a track
reconstruction processor used in the CDF experiment at Tevatron
accelerator. It reconstructs tracks in about 20 $\mu s$ in two steps:
first, low resolution tracks (\textit{roads}) are found in each event
among the energy deposits left in the tracking detector by charged
particles; second, track fitting is performed on all possible
combinations of hits inside a road.  This algorithm uses a linearized
approximation to track-fitting as implemented in hardware (described
in greater detail in~\cite{SVT3}).  With the linearized track fit of
the SVT approach, the determination of the track parameters ($p_i$) is
reduced to a simple scalar product:
\[
p_i = \vec{f_i} \cdot \vec{x_i} + q_i,
\]
where $\vec{x_i}$ are input silicon hits, and $\vec{f_i}$ and $q_i$ are 
pre-defined constant sets. For each set of hits, the algorithm
computes the impact parameter $d_0$, the azimuthal angle $\phi$, 
the transverse momentum $p_\mathrm{T}$ , and the $\chi^2$ of the
fitted track by using simple operations such as memory lookup and 
integer addition and multiplication.

We ported the track fitting as it is well suited to parallelization -
each track can be handled independently.

\subsection{Code implementation}
The starting point of our studies is the SVT track fitting
simulation code, written in the \textsc{c} language. SVT track fitting
is divided into three main functions: first, the unpacking of input
data and filling of all the necessary data structures; second, the
computation of all possible combinations of hits in each road and
third, the linearized track fit of each combination of hits. Three
main loops are present - on events, roads and hit combinations.

To be run on \textsc{nvidia} GPUs, the code has been ported to \textsc{cuda}:
each step -- unpack, combine and track fit -- is performed by a
specific kernel; the three nested loops are unrolled so that each GPU
thread processes a single combination of hits.  The \textsc{cuda}
implementation makes use of \textsc{thrust}~\cite{bib_thrust}, a C++
template library for \textsc{cuda}, in the unpacking step. The
existence of template libraries such as \textsc{thrust} is an
advantage of the \textsc{cuda} environment.

To implement the algorithm to run on an \textsc{amd} GPU, we have ported the
\textit{combine} and \textit{track fit} \textsc{cuda} kernels to
\textsc{opencl}, which requires minimal changes. Because the
\textsc{thrust} template libraries can only be used with
\textsc{cuda}, we resort to unpacking serially on the CPU.

To run on MIC, where cores are more powerful but fewer in number, we
adopted the so-called \textit{embarrassingly parallel} approach and
used \textsc{pragma OpenMP} for statements to unroll only the external
loop on the events, so that each core processes a single event: the
porting requires much less effort compared to \textsc{cuda}, but the
level of parallelism is limited.

\section{Experimental setup and data flow}
The many-core devices used in this study are listed in
Table~\ref{tab_hwspecs}. The GPUs include a less expensive gaming
class GPUs (the \textsc{nvidia} GTX and \textsc{amd} Radeon cards) and ones optimized
for scientific computing (Tesla). The MIC corresponds to a Xeon Phi
introduced in November 2012.

\begin{table*}[tbp]
  \caption{Capabilities of the many-core devices used in this study,
    according to the manufacturer's specifications. The first three
    are \textsc{nvidia} GPUs, the MIC 5110P is an Intel Xeon Phi, and
    the final one is an \textsc{amd} GPU.  For Xeon Phi, the ``cores'' column
    counts the HW threads per core as equivalent to a GPU core.}
  \label{tab_hwspecs}
  \centering
  \begin{tabular}{|l|c|c|c|c|c|}
    \hline
    Model & Tesla M2050 & Tesla K20m & GeForce GTX  Titan & MIC 5110P & Radeon HD 7970 \\
    \hline
     \hline
    Performance (SP, GFlops) & 1030 & 3520 & 4500 & 2022 & 3790\\
    Memory bandwidth  (GB/s) & 148 & 208 & 288  & 320 & 264\\   
    Memory size (GB) & 3 & 5 & 6 & 8 & 3\\
    Number of cores & 448 & 2496 & 2688 & 240 & 2048\\
    Clock speed (GHz) & 1.15 & 0.706 & 0.837 & 1.053 & 1.375\\
    \hline
  \end{tabular}
\end{table*}

\begin{figure}[tbp]
\centering
\includegraphics[width=3.5in]{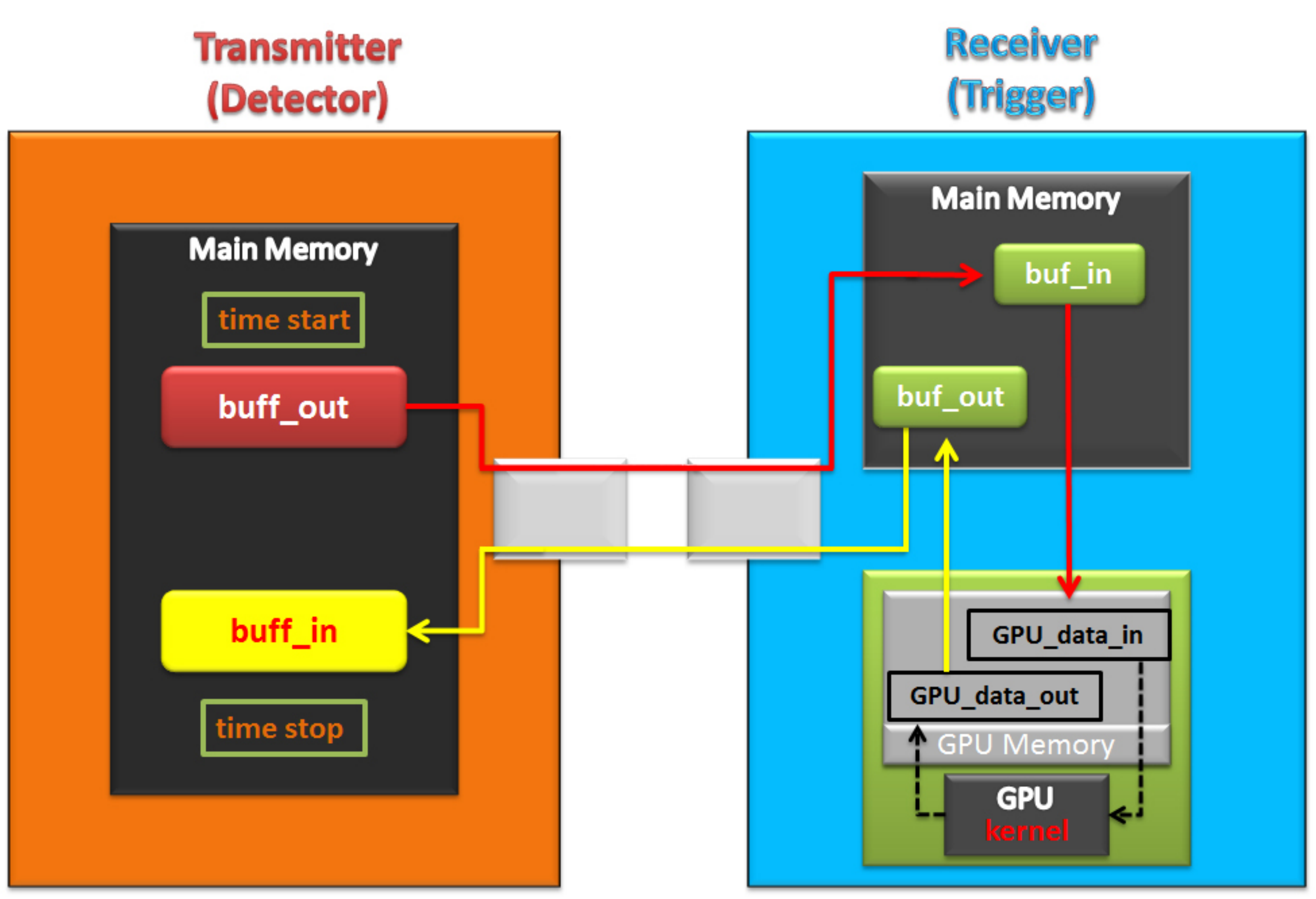}
\caption{Data flow. Data is sent from the transmitter PC to the
  receiver PC, where it is processed by the GPU before being returned
  to the transmitter PC. The transmitter plays the role of the
  detector as the source of the data and as an upstream trigger
  processor as the data's ultimate sink. The receiver PC plays the
  role of a component in the trigger system. }
\label{fig:data_flow}
\end{figure}

To measure the data transfer latency we use a computing cluster composed of 
12 identical nodes.  Each node contains a Intel Xeon E6520 2.4 GHz 
CPU and two Tesla M2075 GPU cards. The nodes are connected by InfiniBand 
communication links using Connect-X2 Mellanox or APEnet+ adapters. 
APEnet+ is an FPGA-based PCIe board supporting peer-to-peer communication with 
Tesla and Kepler cards~\cite{apenet2010}.
Two nodes of this cluster are used to measure data transfer latency,
one acting as a transmitter and the
other as a receiver.  Data are transferred from the transmitter to the 
receiver, processed on the GPU and sent back to the receiver (see
Fig.~\ref{fig:data_flow}).  
The latency for a complete loop is measured on the transmitter using 
standard \textsc{c} libraries.
In this setup, the transmitter can represent the detector, as
the source of the data, or an upstream trigger processor, as
the ultimate sink of the data, while the receiver is the
trigger system: the time to transfer data to the receiver is thus a
rough estimate of the latency to transfer the data from the detector
front-end to the trigger system.

We have an additional setup for testing the \textsc{opencl}
implementation of the track fitting algorithm. Here, we use a 3.07 GHz
Intel Core i7 CPU 950, which has four cores and up to eight
computation threads. We run the algorithm serially on this CPU (one
core), and also run a CPU-based \textsc{opencl} algorithm which makes
use of the multi-core architecture. Additionally, we have an \textsc{amd}
Radeon HD 7970 GPU in a PCIe slot in this setup, on which we also run
the \textsc{opencl} algorithm.

\section{Results}
The input data consists of events with a fixed number of roads and
combinations: each event has 2048 combinations to be fitted. To
explore different data-taking conditions, the number of events ranges
from one to 3000, \emph{i.e.,} between 2048 to about six millions of
combinations to fit.
 
\subsection{Data processing}
\begin{figure}[!t]
  \centering
  \includegraphics[width=0.9\linewidth]{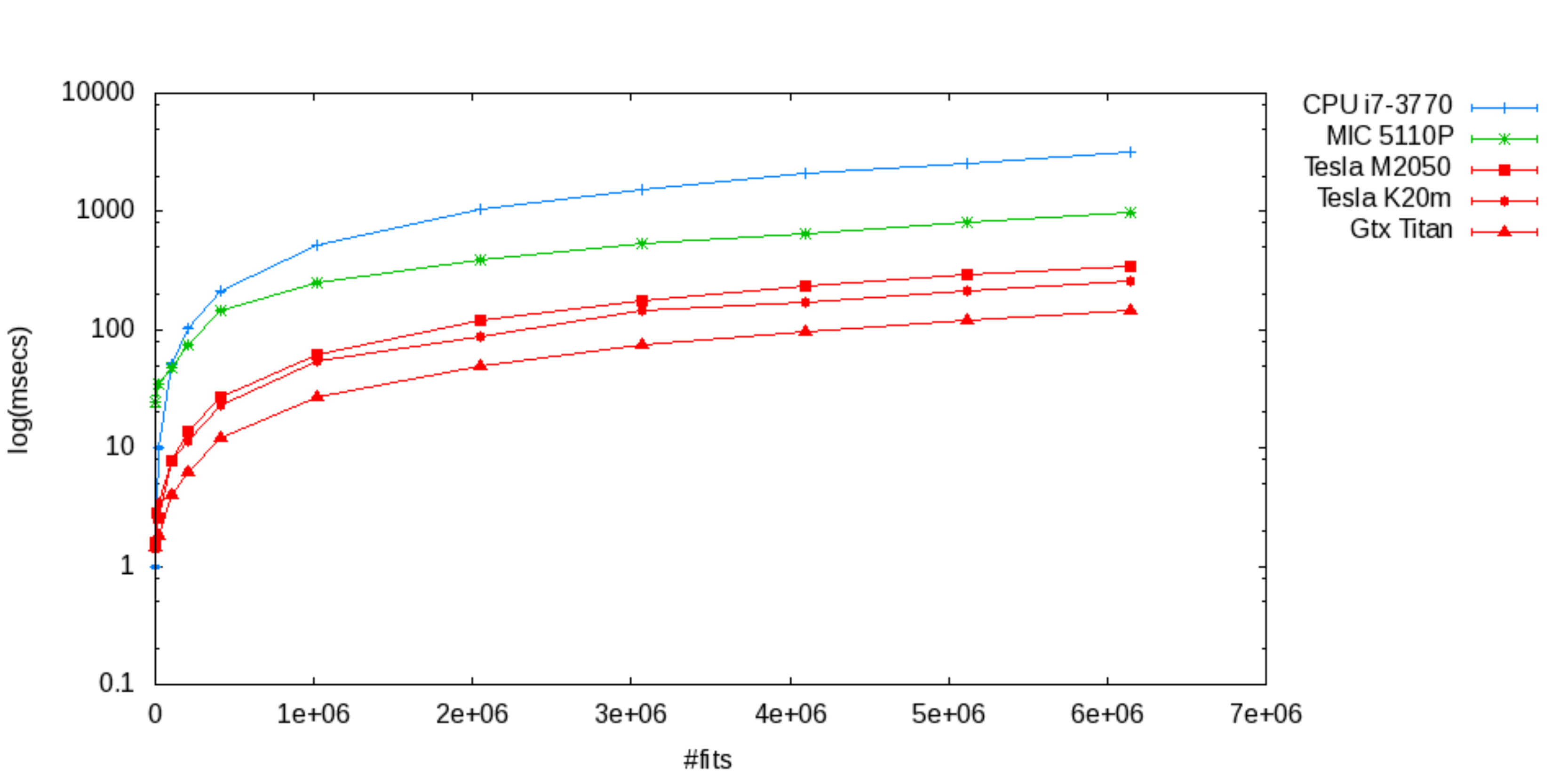}
  \caption{Algorithm-only comparison for timing as a function of the
    number of track fits. We compare timing on CPUs (serial), Intel
    MIC (embarrassingly parallel), and GPUs (fully parallel), in
    blue, green, and red, respectively. The GPUs exhibit the best
    performance due to the full parallelization. }
  \label{fig:algo_only_timing}
\end{figure}

Each data sample is processed 100 times by the track fitting algorithm. 
The average latency as a function of the number of fits is presented in 
Fig.~\ref{fig:algo_only_timing} for the serial,
embarrassingly parallel and parallel algorithms. We see that the
embarrassingly parallel algorithm gives a modest increase with respect
to the serial (CPU) algorithm. Switching to a fully parallel algorithm
affords a much more significant speed improvement. 
\begin{figure}[!t]
  \centering
  \includegraphics[width=0.9\linewidth]{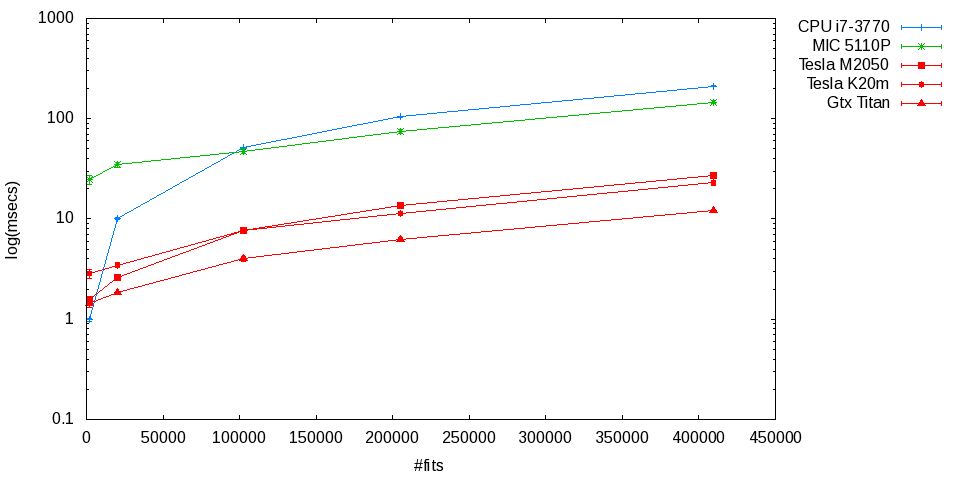} 
  \caption{Algorithm-only comparison for timing as a function of the
    number of track fits: zoom in the low number of fits region. At
    low number of fits, the CPU performs better, due to start-up
    costs associated with data transfers to the accelerator card.}
  \label{fig:algo_only_timing_zoom}
\end{figure}
\begin{figure}[!t]
  \centering
  \includegraphics[width=0.9\linewidth]{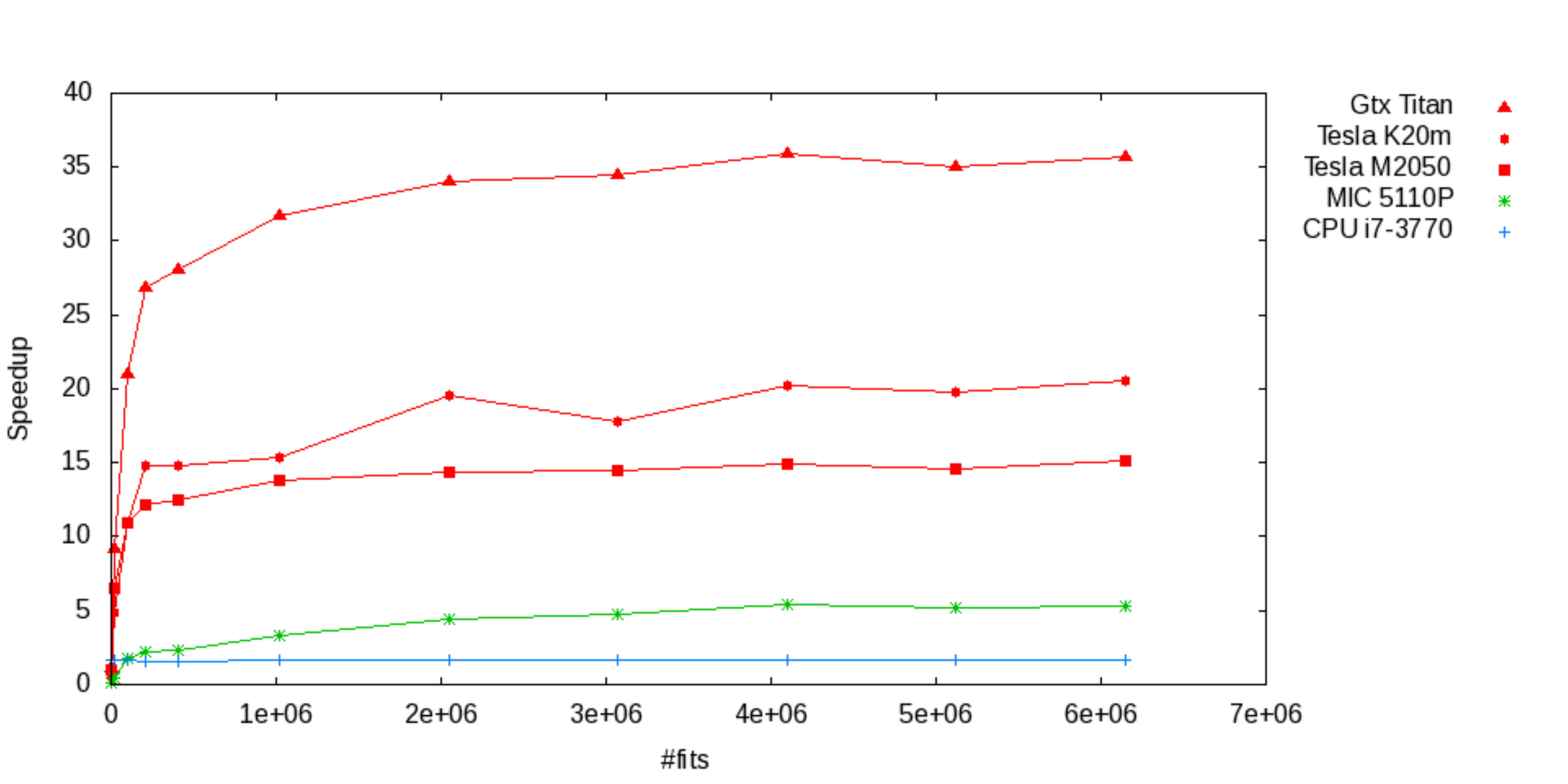}
  \caption{Speed-up with respect to a standard CPU (Intel Xeon
    E5630). The speed-ups plateau after about two million fits.}
  \label{fig:algo_only_speedup}
\end{figure}
The accelerator card's performance drop with decreasing number of fits, as can be seen 
in Fig.~\ref{fig:algo_only_timing_zoom}, due to overhead.
Figure~\ref{fig:algo_only_speedup} shows the speed-up with respect to the serial 
algorithm run on a standard CPU (Intel Xeon E5630): the maximum gain is 
obtained processing at least 500 events. This means that to fully exploit 
parallel architectures millions of fits have to be performed in parallel.  

%

\subsubsection{Breakdown of computing time}
\begin{figure*}[!t]
\centering
\subfigure[]
{\label{fig:breakdown_MIC}
\includegraphics[width=0.45\linewidth]{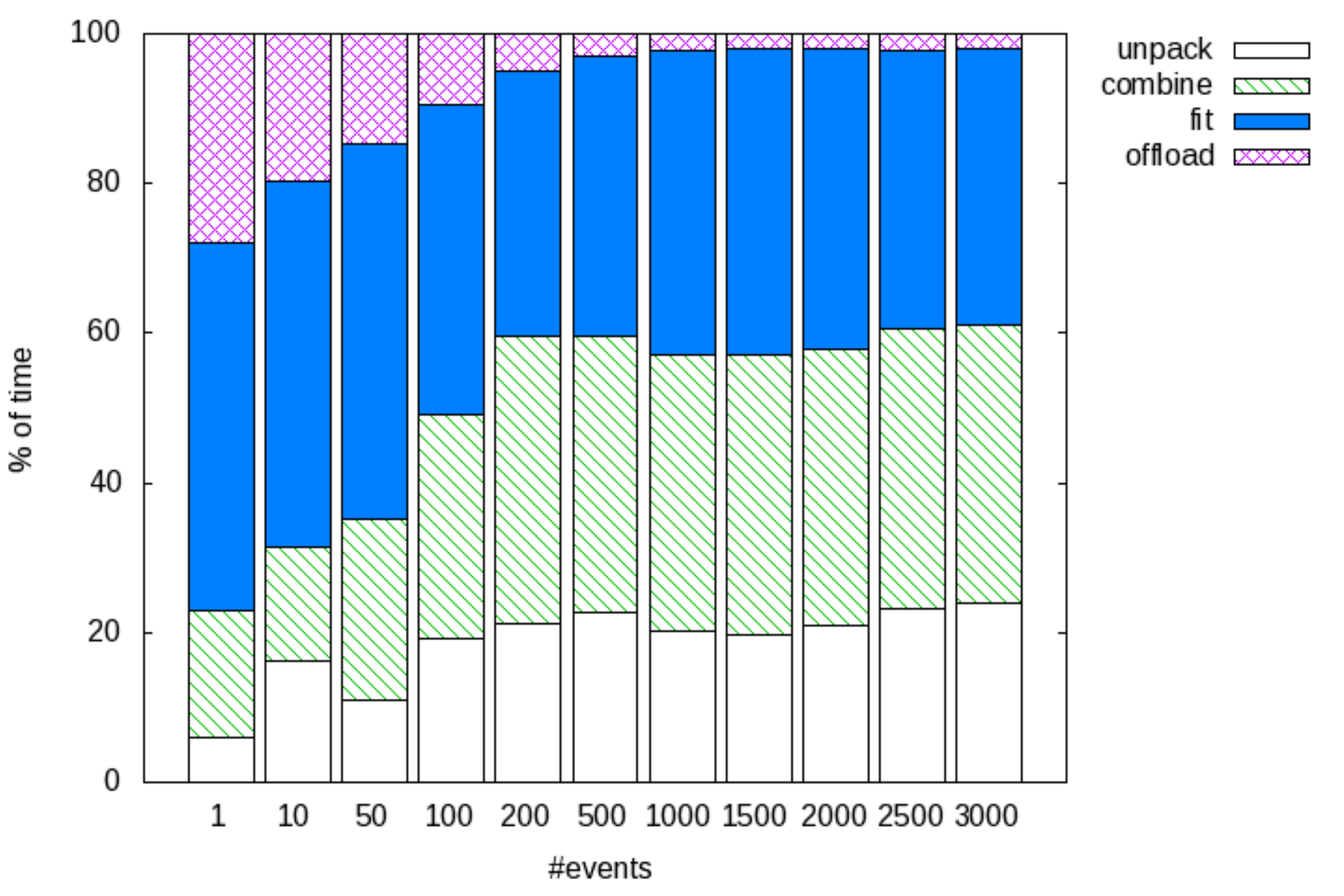}}
\subfigure[]
{\label{fig:breakdown_GPU}
\includegraphics[width=0.45\linewidth]{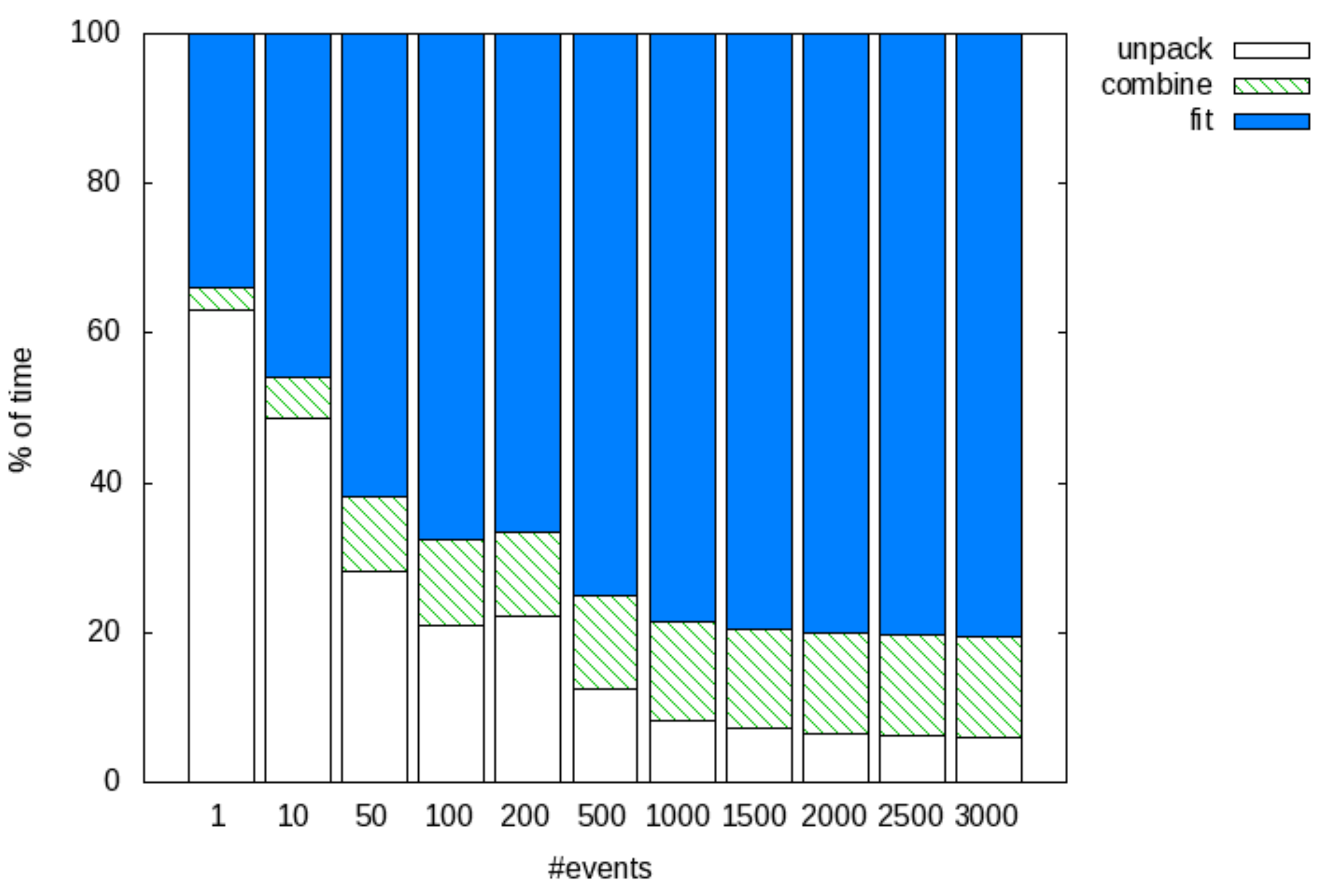}}
\caption{Breakdown of computing time for MIC (a) and the GTX Titan GPU
  (b). White corresponds to unpacking, green hash corresponds to
  generating hit combinations, solid blue is the linearized track fit,
  and magenta cross-hatch corresponds to offloading (MIC only). For
  MIC, combinations and fitting take the same amount of time for large
  number of events. For GPU, fitting dominates. }
\label{fig:breakdown}
\end{figure*}

In Fig.~\ref{fig:breakdown} we show the fractional time spent in
various parts of the algorithm for the embarrassingly parallel algorithm 
(on Intel MIC) and the parallel algorithm (on \textsc{nvidia} Titan GPU), as a 
function of the number of fits. On both accelerator cards the fractional times
are constant for more than 500 input events, where computing resources are saturated. 
Unlike the MIC, the fit stage takes most of the time on the GPU: this 
could be caused by the intense memory access frequency intrinsic to
this part of the algorithm.  

\subsection{Data processing in \textsc{opencl}}
\begin{figure}[!t]
  \centering
  \includegraphics[width=0.9\linewidth]{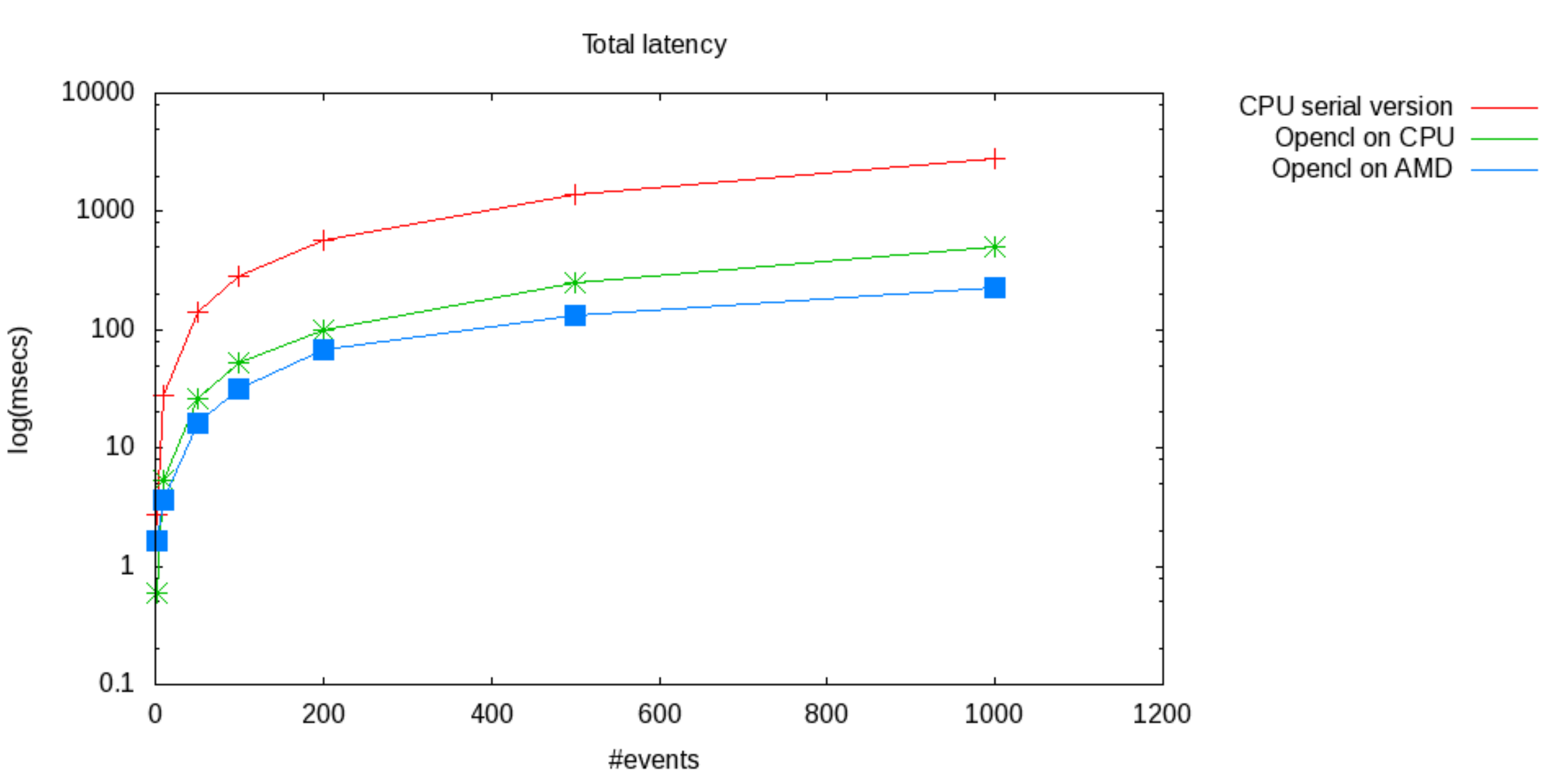}
  \caption{The timing of the \textsc{opencl} algorithm as a function of the
    number of track fits. We compare timing running the algorithm serially on a CPU, 
    using \textsc{opencl} on the CPU, and on an \textsc{amd} GPU.}
  \label{fig:opencl_timing}
\end{figure}

We also measure the track fitting algorithm latency in an
implementation using \textsc{opencl}. The \textsc{opencl} tools have
the advantage of being an open standard, while the \textsc{cuda} tools
are only compatible with \textsc{nvidia} GPUs. Additionally, \textsc{opencl}
can work also on CPUs and on the Xeon Phi. For our tests, due to
limitations on the available space for storing data on the GPU, we
only test the algorithm on up to 1000 events (\textit{i.e.} about 2
million combinations to fit). In Fig.~\ref{fig:opencl_timing} we show
the results of algorithm latency measurements for three different
modes: running the serial algorithm on the CPU (single-core), running
the \textsc{opencl} algorithm on the CPU (multi-core), and running the
\textsc{opencl} algorithm on the \textsc{amd} Radeon GPU. The \textsc{opencl}
implementation of the algorithm on the CPU provides a significant
speedup---about a factor of 5---over running the algorithm serially on
the CPU. Running the same \textsc{opencl} algorithm on the GPU
provides an even greater speedup due to the increased number of cores
and parallel threads that can be run, though there is additional
latency to copy data into and out of the GPU that makes running on the
GPU take longer for small numbers of combinations to fit.
Surprisingly, though \textsc{opencl} is an open standard, we find that
we are unable to run the same \textsc{opencl} code on \textsc{nvidia} GPUs. On two
different installations with two different video cards, we find
incorrect results running the code that ran successfully on the \textsc{amd}
card. At present it is unclear what the root cause  is.

\subsection{Data transfer}
\begin{figure*}[!t]
\centering
\subfigure[]
{\label{fig:standardDT}
\includegraphics[width=0.3\linewidth]{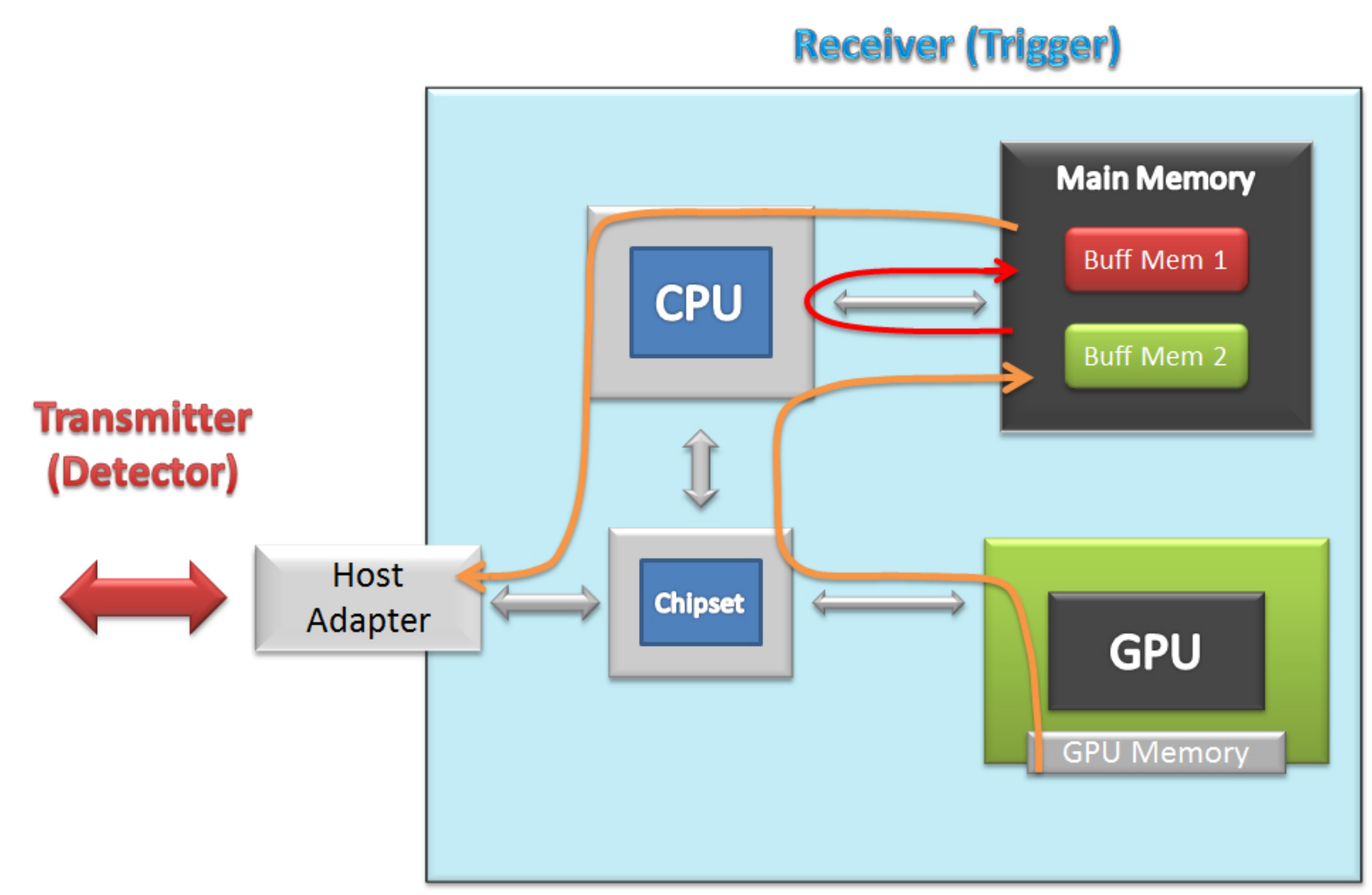}}
\hspace{1mm}
\subfigure[]
{\label{fig:GPUDirectV1}
\includegraphics[width=0.3\linewidth]{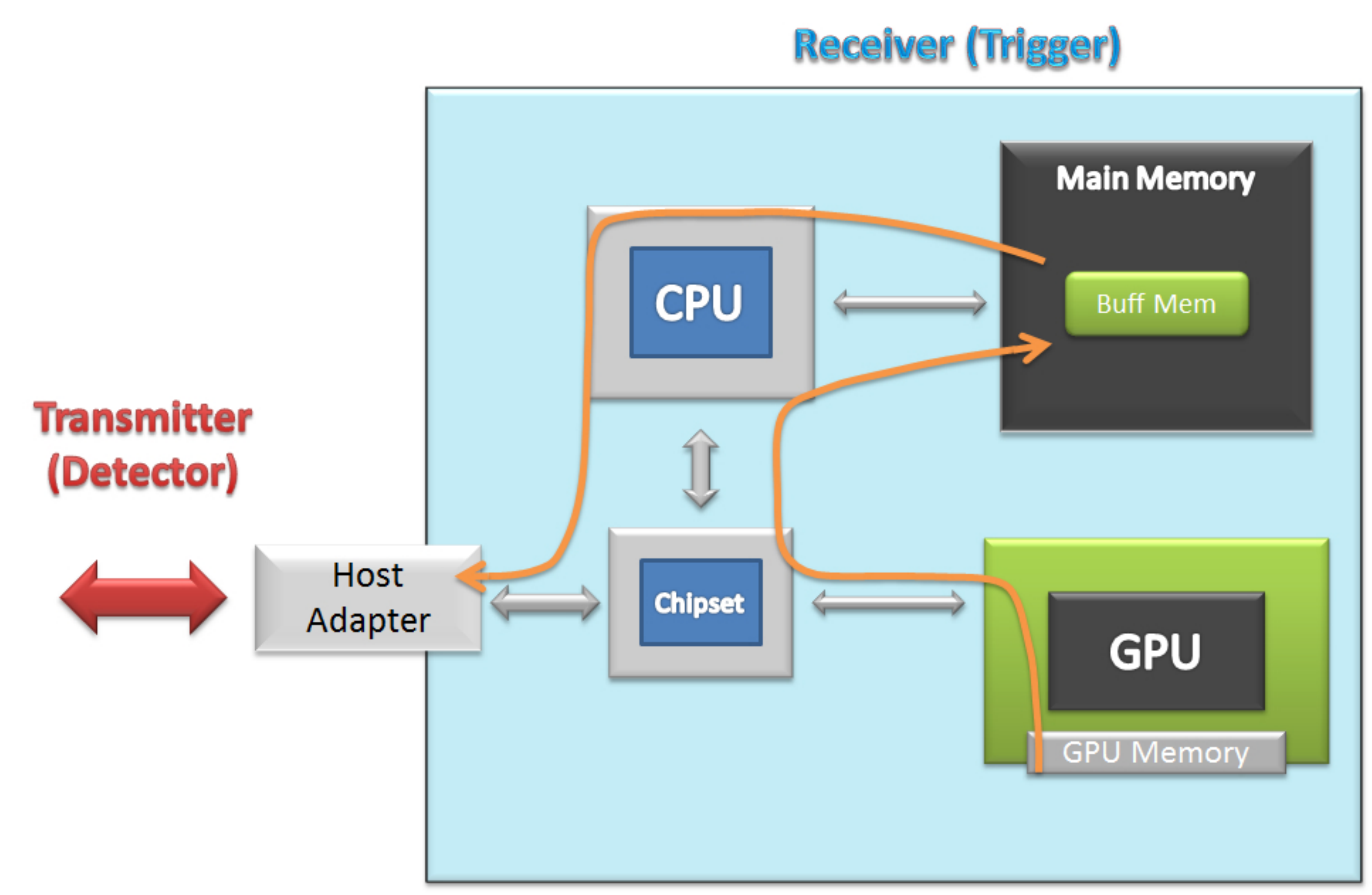}}
\subfigure[]
{\label{fig:GPUDirectV2}
\includegraphics[width=0.3\linewidth]{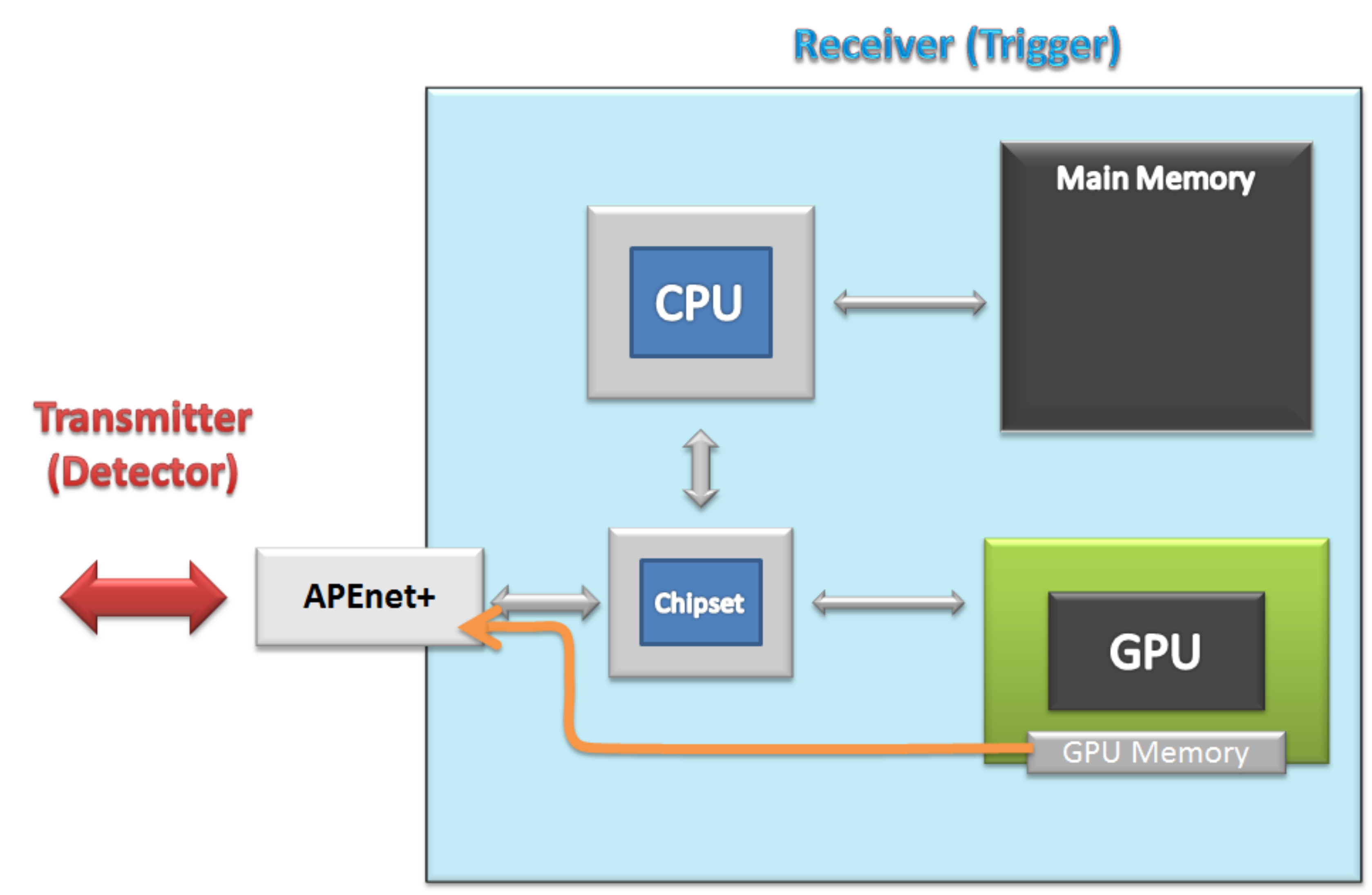}}
\caption{Standard data transfer (a), via GPUDirect (b) and via
  GPUDirect with P2P support (c). In (a), two buffers are required in
  the main memory. In GPUDirect (b), one of the main memory buffers is
  eliminated. In GPUDirect with P2P support, data is sent directly
  from the APEnet+ transceiver to the GPU memory.}
\end{figure*}

The experimental setup described in Fig.~\ref{fig:data_flow} allows us to 
test different data transfer  strategies to the GPU. The standard data transfer 
strategy is via the system memory, where the 
PCIe adapter card and the GPU allocate \emph{separate} buffers on the
system memory for the copy (as shown in Fig.~\ref{fig:standardDT}).
This is inefficient, as the data are copied twice in the 
system memory before being transferred to the GPU/PCIe card. 
Data may also be transferred using Direct Memory Access (DMA, 
GPUDirect~\cite{bib_GPUDirect}) to the CPU memory:
the PCIe card and the GPU share the \textit{same} buffer on the CPU memory; as 
a result the data are copied only once in the CPU memory 
(Fig.~\ref{fig:GPUDirectV1}). 
With our experimental setup two additional copy strategies can be tested
which are the results of different levels of 
optimization of the GPUDirect protocol:
\begin{itemize}
\item \textsc{cuda}-Aware MPI, where the copy latency is further reduced
by automatically allocating the buffer on the CPU memory;
\item peer-to-peer (P2P) strategy, when data are transferred 
directly to the GPU, without any  
intermediate copy to the CPU (Fig.~\ref{fig:GPUDirectV2}).
\end{itemize}

\begin{figure}[!t]
  \centering
  \includegraphics[width=0.85\linewidth]{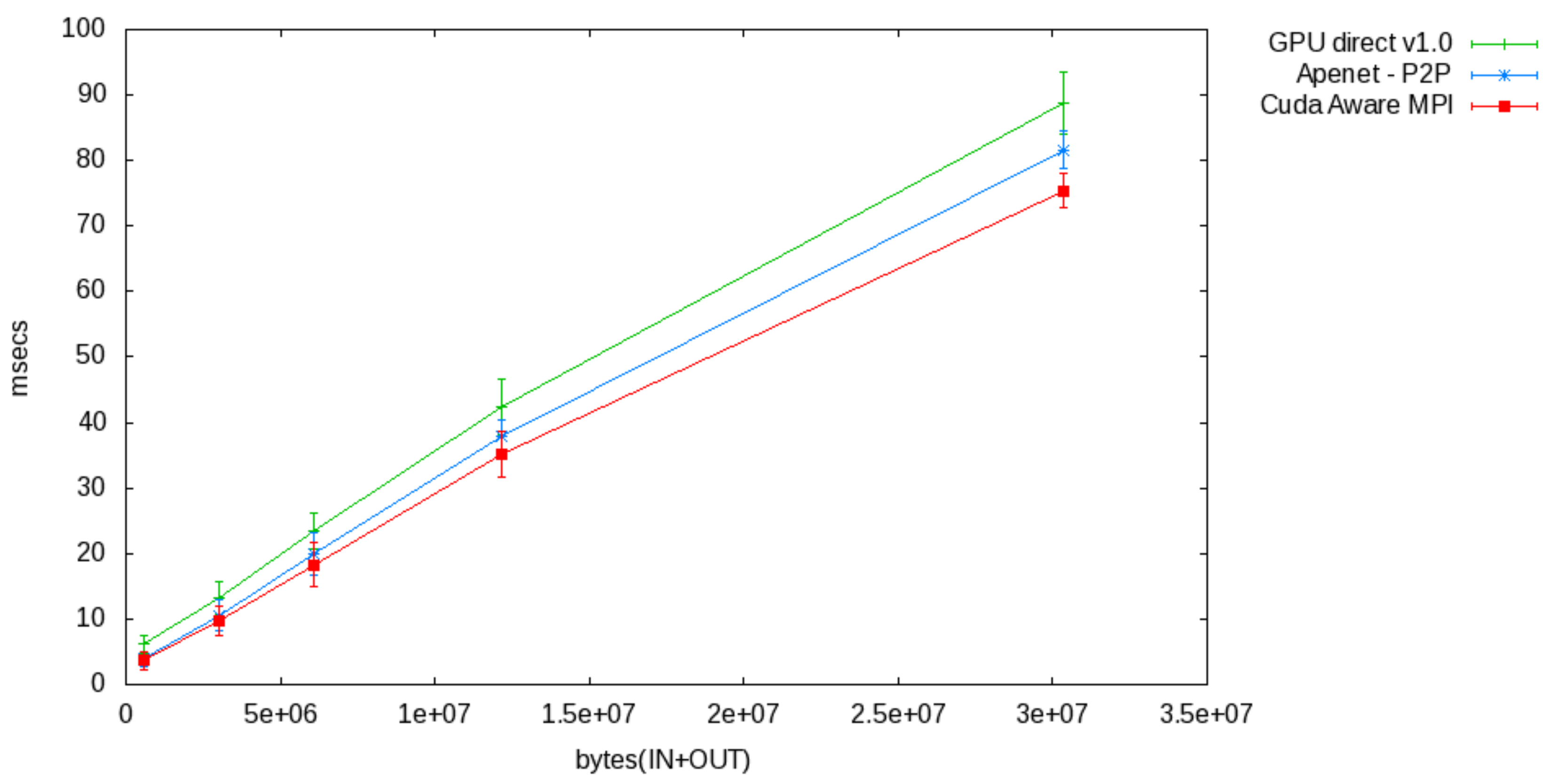}
  \caption{Total latency (data transfer, copy to/from the GPU and data
    processing on the GPU) as a function of data buffer sizes, for
    three different levels of optimization of GPUDirect: v1.0,
    \textsc{cuda}-aware MPI and P2P. The smallest transferred data
    packed is 600 kB.  \textsc{cuda}-aware MPI shows the best
    performance for larger packet size.}
  \label{fig:xferlatency}
\end{figure}

\begin{figure}[!t]
  \centering
  \includegraphics[width=0.85\linewidth]{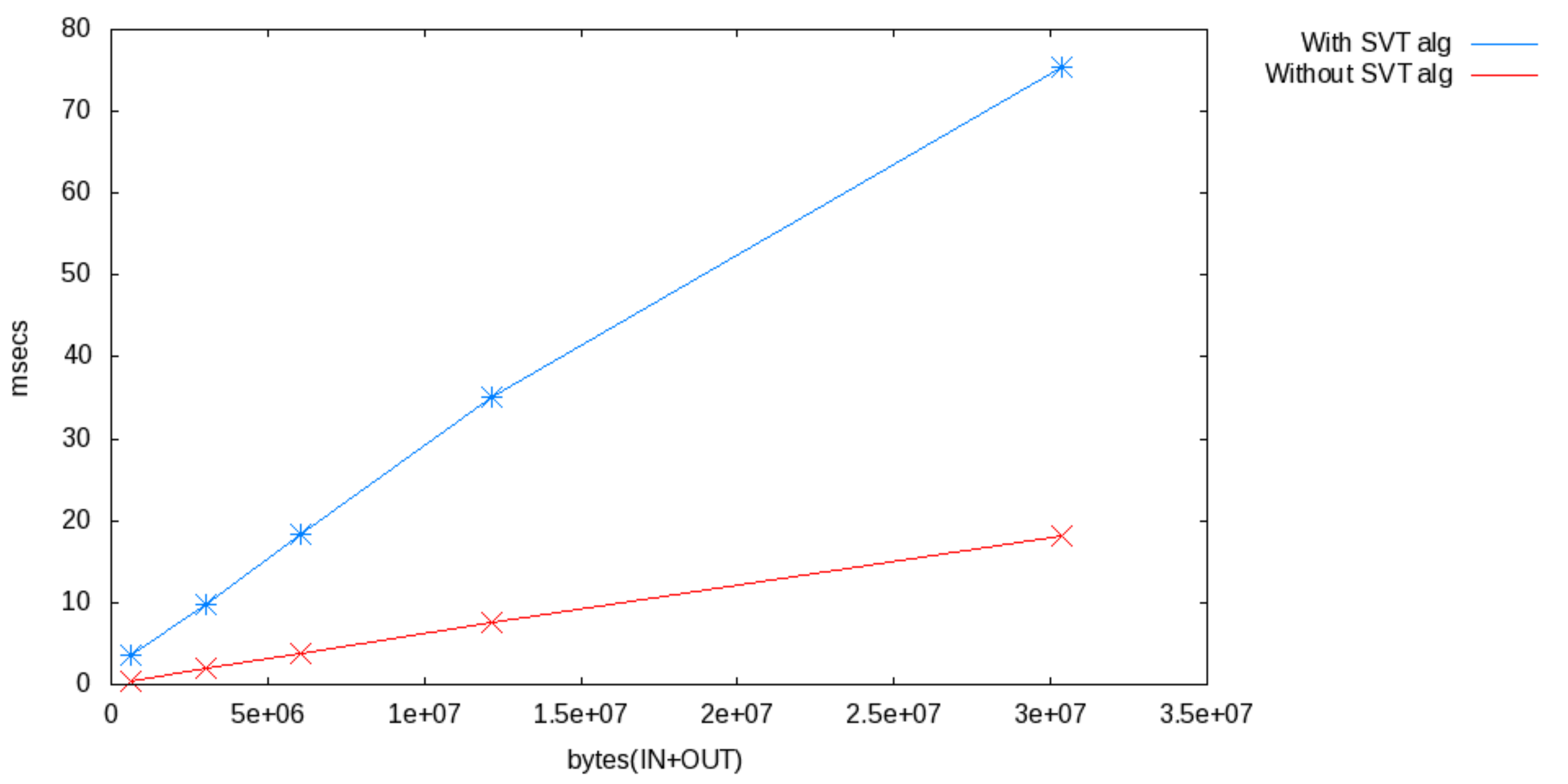}
  \caption{Time per fit, in msec. The two curves show total timing
    with and without calculations performed on the GPU, thereby
    showing the considerable time spent in data transfer. About
    20-25\% of the time is spent in data transfer.}
  \label{fig:transferOnly}
\end{figure}

In Fig.~\ref{fig:xferlatency} we show the total latency (data
transfer, copy to/from the GPU and data processing on the GPU) as a
function of data packet size when data are transferred using GPUDirect
v1.0, \textsc{cuda}-aware MPI and P2P. For the packet sizes considered in this test 
 \textsc{cuda}-aware MPI gives the best performance. This is expected as 
P2P is optimized for small packet sizes
 (see also~\cite{NSS2012} and~\cite{bib_mvapich}). As a matter of fact, for larger packet size, the channel 
throughput becomes dominant: the shortest transfer 
time of CUDA aware-MPI system is easily explained comparing the link bandwidth of Mellanox board (40 Gb/s) 
with the smaller throughput of a APEnet+ single link (30 Gb/s).  
The data transfer latency accounts for a significant part of the total
latency, as can be seen in Fig.~\ref{fig:transferOnly}: about 20-25\%
of total latency is due to moving the data to and from the GPU.


\section{Conclusions}
We have implemented a full version of the CDF SVT tracking algorithm
on GPUs and Intel MIC. We examined a staged approach to using
accelerator cards in a hadron collider trigger application and
compared additional software tool sets. We have demonstrated that in
this application, significant gains can be achieved with the
`embarrassingly parallel' approach on an Intel MIC architecture, with
the smallest amount of required changes to an existing serial code
base. Additionally, we have implemented a parallelized algorithm using
\textsc{cuda} and \textsc{opencl}. Better performance is achieved with
GPUs and a more complete event-level parallelization using these
tools.  We have run the parallelized algorithm on a multi-core CPU and
GPU, showing a boost in performance over serial CPU computation for
even small numbers of events.  We have also updated latency studies
and shown that for larger packet sizes (greater than 600 kB),
\textsc{cuda}-aware MPI outperforms P2P.  Even at large packet size,
the data transfer takes an appreciable fraction of the total algorithm
time (about 20-25\%).

%

\section*{Acknowledgment}
The authors would like to thank the Fermilab staff,  the FTK group at the 
University of Chicago and the INFN-APE group in Rome for their support. This work was supported by the
U.S. Department of Energy, the U.S. National Science Foundation and the Italian
Istituto Nazionale di Fisica Nucleare. This work was partially supported by the 
EU Framework Programme 7 project EURETILE under grant number 247846. 

\bibliographystyle{IEEEtran}

\bibliography{gpu}

\end{document}